\documentclass[12pt,a4paper]{article} 
\usepackage{epsfig}

\newcommand{\beq}{\begin{equation}}   
\newcommand{\eeq}{\end{equation}}   
\newcommand{\beqn}{\begin{eqnarray}}  
\newcommand{\eeqn}{\end{eqnarray}}   
\newcommand{\nn}{\nonumber}

\usepackage{chir_layout}   

\def\desy{c}
\def\ors{a}
\def\rmii{b}
\def\rmiii{d}
\def\infn{e}
\def\rmi{f}
\def\liv{g}
\def\hum{h}

\begin{document}  
% \maketitle
\begin{titlepage}
{
% \vspace{-2.5cm}
\normalsize
\begin{flushright}
\hfill %\parbox{100mm}
{DESY 09-112,~LPT-Orsay/09-69,~LTH835,\\~HU-EP-09/42,~RM3-TH/09-16,~ROM2F/2009/13\\ }
\end{flushright}
}%[10mm]
% \vspace{-0.5cm}
\begin{center}
  \begin{Large}
    \textbf{A proposal for $B$-physics on current lattices \\
           \unboldmath} 
  \end{Large}
\end{center}

%\vspace{-0.5cm}
\begin{figure}[h]
  \begin{center}
  \includegraphics[draft=false]{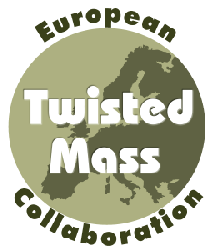}
  \end{center}
\end{figure}

% \vspace{-0.5cm}
\baselineskip 20pt plus 2pt minus 2pt
\begin{center}
  \textbf{
    B.~Blossier$^{(\ors)}$,
    P.~Dimopoulos$^{(\rmii)}$,
    R.~Frezzotti$^{(\rmii)}$,
    G.~Herdoiza$^{(\desy)}$,\\
K.~Jansen$^{(\desy)}$,~V.~Lubicz$^{(\rmiii,\infn)}$,~G.~Martinelli$^{(\rmi)}$,~C.~Michael$^{(\liv)}$,~G.C.~Rossi$^{(\rmii)}$,\\
    A.~Shindler$^{(\liv)}$,
    S.~Simula$^{(\infn)}$,
    C.~Tarantino$^{(\rmiii,\infn)}$,
    C.~Urbach$^{(\hum)}$}
\end{center}

\begin{center}
  \begin{footnotesize}
    \noindent 

$^{(\ors)}$ Laboratoire de Physique Th\'eorique (B\^at.~210), Universit\'e de
Paris XI,\\ Centre d'Orsay, 91405 Orsay-Cedex, France
\vspace{0.2cm}

$^{(\rmii)}$ Dip. di Fisica, Universit{\`a} di Roma Tor Vergata and INFN, Sez.
di Roma Tor Vergata,\\ Via della Ricerca Scientifica, I-00133 Roma, Italy
\vspace{0.2cm}

$^{(\desy)}$ NIC, DESY, Platanenallee 6, D-15738 Zeuthen, Germany
\vspace{0.2cm}

$^{(\rmiii)}$ Dip. di Fisica, Universit{\`a} di Roma Tre, Via della Vasca Navale
84, I-00146 Roma, Italy
\vspace{0.2cm}

$^{(\infn)}$ INFN, Sez. di Roma III, Via della Vasca Navale 84, I-00146 Roma, Italy
\vspace{0.2cm}

$^{(\rmi)}$ Dip. di Fisica, Universit{\`a} di Roma La Sapienza and INFN, Sez. di Roma La Sapienza,\\
P.le A. Moro 5, I-00185 Roma, Italy
\vspace{0.2cm}

$^{(\liv)}$ Theoretical Physics Division, Dept. of Mathematical Sciences,
\\University of Liverpool, Liverpool L69 7ZL, UK
\vspace{0.2cm}

$^{(\hum)}$ Institut f\"ur Elementarteilchenphysik, Fachbereich Physik,\\ 
Humbolt Universit\"at zu Berlin, D-12489, Berlin, Germany
\vspace{0.2cm}

\vspace{0.2cm}

\end{footnotesize}
\end{center}
\end{titlepage}

\begin{abstract}{A method to extract $B$-physics parameters ($b$-quark mass and $f_B$,  
$f_{B_s}$ decay constants) from currently available lattice data is presented and tested. The
approach is based on the idea of constructing appropriate ratios of heavy-light meson masses 
and decay constants, respectively, possessing a precisely known static limit, and evaluating them 
at various pairs of heavy quark masses around the charm. Via a smooth interpolation in the 
heavy quark mass from the easily accessible charm region to the asymptotic point, $B$-physics
parameters are computed with a few percent (statistical + systematic) error using recently
produced $N_f=2$ maximally twisted Wilson fermions data.}
\end{abstract}

\section{Introduction}
\label{sec:INTRO}

Heavy flavour physics is a corner of the Standard Model where chances are higher to uncover 
signals of new physics~\cite{REVSM,UT}. However, to extract from experiments useful phenomenological 
information, it is mandatory to have an accurate knowledge of the relevant hadronic matrix elements of the 
effective weak Hamiltonian. For low mass states (up to around the charm mass) lattice QCD (LQCD) represents the ideal 
framework where such calculations can be performed with well under control systematic errors~\cite{REVLAT}. 

Due to present day computer limitations, it is not possible, however, to work directly with the heaviest quarks (as the 
$b$-quark) propagating on the simulated lattice. Various strategies, more or less inspired to the heavy quark effective 
theory (HQET)~\cite{HQET,HQET1}, have been devised to circumvent this intrinsic difficulty, which go from 
non-perturbative matching of HQET onto QCD~\cite{SOMMER} to finite size scaling methods with relativistic heavy 
quark(s)~\cite{FSSM}. Relativistic heavy-quark actions designed (highly tuned) to have reduced cutoff effects~\cite{FERMILAB} 
have also been employed for this purpose. Encouraging results have been obtained by several groups~\cite{GAMIZLAT08} 
though different is the level at which the various relevant systematic effects are controlled.

In this work we wish to present a novel approach to $B$-physics in which the $b$-mass point is attained by
interpolating from the charm region to the asymptotic infinite mass regime suitable ratios of heavy-light 
($h\ell$) meson masses and decay constants, computed at a number of pairs of quark mass values lying 
slightly below and somewhat above the charm mass. The key feature of the approach is the use of ratios of 
physical quantities which by construction have a well defined and exactly known infinite $h$-quark mass 
limit. Injecting knowledge of meson masses and decay constants at the charm region, their $h$-quark mass 
evolution can be computed by a chain of successive steps  
up to values as large as about twice the charm mass. The $b$-physics region is finally reached through 
an interpolation from the simulated points to the exactly known infinite $h$-quark mass value.

A first test of the viability of the method is presented here. It has been carried out by exploiting the 
unquenched $N_f=2$ data recently produced by the ETM Collaboration~\cite{ETMCD,VCS} which makes 
use of maximally twisted Wilson fermions~\cite{TM}. The results obtained in this feasibility study are 
very encouraging and compare nicely with the unquenched determinations today available in the 
literature~\cite{LITERUQ} as well as with PDG numbers~\cite{REVSM}. We get 
\beqn
&&\hat{\mu}^{\overline{MS},N_f=2}_b(\hat{\mu}^{\overline{MS},N_f=2}_b)=4.63(27)~{\mbox{GeV}}\, ,
\label{UQRES1}\\
&&f_B=194(16)~{\mbox{MeV}}\, ,\label{UQRES2}\\        
&&f_{B_s}=235(12)~{\mbox{MeV}}\, ,\label{UQRES3}    
\eeqn
where, as indicated explicitly in eq.~(\ref{UQRES1}), the $b$-mass has been run in a world with two 
(active) flavours. The results in eqs.~(\ref{UQRES1}) to~(\ref{UQRES3}) represent a ``first principle''
determinations of $B$-physics parameters with errors whose magnitude
can be systematically reduced. The quoted uncertainty will be discussed in sects. 2 and 3
for eqs.~(\ref{UQRES1}) and (\ref{UQRES2})-(\ref{UQRES3}), respectively.

A few observations are in order here. First of all we would like to remark that the results above are 
extracted from unquenched LQCD data where $u$ and $d$ light fermions are dynamical, while 
heavier quarks are introduced only as valence quarks. This scheme is what goes under the name of ``partially 
quenched'' setting (see ref.~\cite{FR2} for a discussion within the twisted mass regularization of QCD).
Systematic errors due to partial quenching are not included in the figures quoted in eqs.~(\ref{UQRES1}) to~(\ref{UQRES3}).
The second observation is that no complicated renormalization steps are required for the method to work, 
because, as noted above, the necessary inputs are (ratios of) physical quantities ($h\ell$-pseudoscalar 
meson masses or decay constants) evaluated at $h$-quark masses around the charm region which are extracted 
from the existing (large volume) lattice configurations produced for the study of pion physics. 

The central value of $\hat{\mu}^{\overline{MS},N_f=2}_b(\hat{\mu}^{\overline{MS},N_f=2}_b)$ in
eq.~(\ref{UQRES1}) may look somewhat higher (though still compatible within statistical errors) than the available 
phenomenological estimates of the $\overline{MS}$ $b$-quark mass at its own scale, which lie in the range 
4.2-4.3~GeV. However, it is not unlikely that, when the quenching of quarks heavier than $u$ and $d$ will be 
removed with the inclusion of dynamical $s$ and, possibly, $c$ quarks, $m_b$ will receive corrections which one 
can argue will tend to make the quantity $\hat{\mu}^{\overline{MS},N_f=4}_b(\hat{\mu}^{\overline{MS},N_f=4}_b)$ 
somewhat smaller than the number given in eq.~(\ref{UQRES1})~\footnote{Indeed, if we evolve the intermediate 
result $\hat{\mu}_b^{\overline{MS},N_f=2}(2~{\rm GeV}) = 5.35(32)$~GeV from 2~GeV to the $b$-quark 
mass scale by using anomalous dimension and $\beta$-function of the $N_f=4$ (rather than $N_f=2$) theory, the 
value of the $b$-quark mass gets lowered by about 3\% compared to the value we give in~(\ref{UQRES1}).}. 

The results~(\ref{UQRES2}) and~(\ref{UQRES3}), instead, are only affected at a level
of less than 1\% by our present uncertainties in the $b$-quark mass, because the $hu/d$-
and $hs$-meson decay constants happen to have a rather mild dependence on the $h$-quark mass. 

The content of this paper is as follows. In sect.~\ref{sec:METHB} we discuss the theoretical basis 
underlying the strategy that we propose to extract the value of the $b$-quark mass from present day 
LQCD data and we provide a rather accurate determination of it with controlled errors. In sect.~\ref{sec:METHF} 
we extend the method to the determination of the $f_B$ and $f_{B_s}$ decay constants. We conclude 
in sect.~\ref{sec:CONC} with a few words on how to improve the quality of the numbers~(\ref{UQRES1}) 
to~(\ref{UQRES3}), and how to extend the present method to other $h\ell$-physics quantities the large 
$h$-quark mass behaviour of which is known. 
We defer to an Appendix some technical details concerning the way chiral and
continuum extrapolations of $h\ell$ meson masses and decay constants are
performed.

\section{$b$-quark mass}
\label{sec:METHB}

In this section we present a simple strategy aimed at determining the value of the $b$-quark mass through 
a smooth interpolation of suitable ratios of $h\ell$ pseudoscalar lattice meson masses from the well accessible  
charm region to the asymptotic (infinite mass) point where these quantities have an exactly known value. 
Inspired by HQET results we consider the lattice ratios ($a=$ lattice spacing) 
\beqn
&&y^L(x^{(n)},\lambda;\hat\mu_{\ell},a)=\nn\\ 
&&=\frac{M_{h\ell}^L(\hat\mu_h^{(n)};\hat\mu_\ell,a)}
{M_{h\ell}^L(\hat\mu_h^{(n-1)};\hat\mu_\ell,a)} \cdot
\frac{\rho(\log \hat\mu^{(n-1)}_h) \hat\mu^{(n-1)}_h}{\rho(\log \hat\mu^{(n)}_h)\hat\mu^{(n)}_h} 
\, ,\quad n=2,\cdots,N \, .\label{RATM}\eeqn 
In eq.~(\ref{RATM}) and in the following by a ``hat'' we denote quark masses renormalized at 2~GeV in the $\overline{MS}$ 
scheme. By $\hat\mu_\ell$ we indicate the renormalized light quark mass, while $\hat\mu_h^{(n)}>\hat\mu_h^{(n-1)}$ 
are pairs of (renormalized) ``heavy'' valence quark masses lying around (from below to somewhat above) the charm mass.
The function $\rho(\log\hat\mu_h)$ is the factor that ``transforms'' the renormalized $\overline{MS}$ quark mass  
at 2~GeV scale into the so-called ``quark pole mass''. In formulae
\beq
\rho(\log\hat\mu_h) \hat\mu_h=\mu_h^{\rm{pole}}\, .
\label{PMA}
\eeq
In continuum perturbation theory (PT) $\rho$ is known up to N$^3$LL (i.e.\ up to next-to-next-to-next-leading-log) 
order terms included~\cite{CR,POLEtoMSbarMASS}. Finally $N$ is the number of $h$-quark masses at which 
the values of the $h\ell$ pseudoscalar lattice meson masses, $M_{h\ell}^L$, are supposed to have been measured. 

The choice of the form of eq.~(\ref{RATM}) is suggested by the HQET (continuum) asymptotic equation~\cite{HQET,HQET1}
\beq
\lim_{\mu_h^{\rm{pole}}\to \infty} \frac{M_{h\ell}}{\mu_h^{\rm{pole}}}={\mbox{constant}}\neq 0\, .
\label{PM}
\eeq
Although the above constant is known to be 1, its value is not really needed here. 

In order to simplify our subsequent analysis we keep fixed the ratio between two successive values of the 
heavy quark masses in eq.~(\ref{RATM}). Calling it $\lambda>1$, we set 
\beq
\lambda=\frac{\hat\mu^{(n)}_h}{\hat\mu^{(n-1)}_h}=\frac{\mu^{(n)}_h}{\mu^{(n-1)}_h}=
\frac{x^{(n-1)}}{x^{(n)}}\, , \qquad  x^{(n)}=\frac{1}{\hat\mu^{(n)}_h}\, .\label{DEFS}
\eeq
Notice that in $\hat\mu_h^{(n)}/\hat\mu_h^{(n-1)}$ the mass renormalization constant factor, 
$Z_P^{-1}$, cancels out~\footnote{We recall that in maximally twisted LQCD the twisted mass 
renormalizes according to $\hat\mu=Z_P^{-1}\mu$. If standard Wilson fermions were to be employed, 
the quantity $\hat{m}_h = Z_{S^0}^{-1}(m_{0h}-m_{\rm cr}$), 
should be used in place of $\hat\mu_h$ in eq.~(\ref{RATM}).}. 

Ratios of the kind defined in~(\ref{RATM}) are introduced with the idea that they might have a smoother 
chiral ($\mu_\ell\to\mu_{u/d}$, with $\mu_{u/d}$ the light quark mass that yields the 
physical value of the pion mass) and continuum limit than each of the individual factors. Setting  
\beqn
&&y(x^{(n)},\lambda; \hat\mu_{u/d}) \equiv
\lim_{\hat\mu_\ell \to \hat\mu_{u/d}}\,\lim_{a\to 0}\, y^L(x^{(n)},\lambda;\hat\mu_{\ell},a) =\nn\\
&&= \lambda^{-1} \frac{M_{h{u/d}}(1/x^{(n)})}{M_{h{u/d}}(1/\lambda x^{(n)})}
\frac{\rho(\log\lambda x^{(n)})}{\rho(\log x^{(n)})} \, ,
\label{RATLIM}
\eeqn
where we have introduced the (continuum limit) shorthand notation
\[
M_{h{u/d}}(1/x)\equiv M_{hu/d}(1/x,\hat\mu_{u/d})\, ,
\]
we observe that (for all $\lambda>1$)  eqs.~(\ref{PMA}) and~(\ref{PM}) imply the following exact property 
\beq
\lim_{x\to 0} y(x,\lambda;\hat\mu_{u/d})=1\, . \label{RATLEY}
\eeq 

{}From lattice data the function $y(x,\lambda;\hat\mu_{u/d})$ can be determined at certain discrete values of $x$ ($x^{(n)}$, 
$n=2,\cdots,N$). In order to extend our knowledge outside these particular points, while at the same time fully exploiting 
the strong constraining power provided by eq.~(\ref{RATLEY}), we imagine proceeding in the following way. Suppose 
the perturbative expansion of $\rho$ has been computed and resummed up to N$^P$LL order. 
Then we can define a tower of $y$-ratios, $y|_{p}$, $p=0,1,\dots,P+1$, such that 
\beq 
y(x,\lambda;\hat\mu_{u/d})\Big{|}_{p}-1
~\stackrel{x \to 0}{\sim}~ {\rm O}\Big{(}\frac{1}{(\log x)^{p+1}}\Big{)}\, , \label{LIM}
\eeq
provided $\rho$ in eq.~(\ref{RATM}) is correspondingly taken at tree-level in the case of $p=0$, or to N$^{p-1}$LL 
order for $p>0$. Then for sufficiently small values of $x$ we parameterize $y|_{p}$ in the form 
\beq
 y(x,\lambda;\hat\mu_{u/d})\Big{|}_{p}=1+\eta_1(\log x,\lambda;\hat\mu_{u/d})x+
\eta_2(\log x,\lambda;\hat\mu_{u/d}) x^2\, ,\label{ANLYTICY}
\eeq 
where the coefficients $\eta_j(\log x,\lambda;\hat\mu_{u/d})$, $j=1,2$, are $p$-dependent  
(though to lighten the notation we do not display explicitly this dependence in the following), 
smooth functions of $\log x$ which tend to zero as $\lambda \to 1$ and to some fixed constant 
as $x\to 0$~\cite{HQET,HQET1}. 
With the ansatz~(\ref{ANLYTICY}) and at any order where PT results for $\rho$ are available, 
it is not difficult to determine the $\eta_j$ coefficients from lattice data, assuming that their 
$\log x$-dependence can be ignored in the range of masses where the above formulae are used.
 
It is important to remark that the ansatz~(\ref{ANLYTICY}) is based on the same kind of 
assumptions under which HQET is usually employed in the study of heavy quark physics. A posteriori, 
we check  that the best fit values taken by the coefficient functions $\eta_j$ come out of a 
reasonable order of magnitude. Indeed we find that $\eta_1 r_0$ and $\eta_2 r_0^2$ are O(1) 
quantities. 

\subsection{Implementing the method}
\label{sec:LLO}

Let us start considering for concreteness the case where $\rho$ is taken up to LL order and 
subsequently compare the results we get in this way with what one would obtain taking 
$\rho$ at NLL-order or at tree-level (at tree-level $\rho=1$). 

\begin{table}[!ht]
\begin{center}
\begin{tabular}{cccccccccc}
\hline \hline
 $\beta$  &&  $a^{-4}(L^3 \times T)$ && $a\mu_{\ell}~=~a\mu_{sea}$  &&  $a\mu_{s}$ && $a\mu_{h}$  \\
\hline 
%&&&&&&&&& \\
3.80      &&  $24^3 \times 48$&& 0.0060, 0.0080  && 0.0200, 0.0250   &&  0.2700, 0.3100   \\
          &&                  && 0.0110, 0.0165  && 0.0300, 0.0360   &&  0.3550, 0.4350   \\
          &&                  &&                 &&                  &&  \hspace*{-1.4cm}0.5200        \\
\hline 
%&&&&&&&&& \\
%&&&&&&&&& \\
3.90      &&  $24^3 \times 48$&& 0.0040, 0.0064 &&   0.0220, 0.0270 && 0.2500, 0.3200   \\
          &&                  && 0.0085, 0.0100 &&   \hspace*{-1.4cm}0.0320         && 0.3900, 0.4600   \\
          &&                  && \hspace*{-1.4cm}0.0150         &&                  &&                  \\
3.90      &&  $32^3 \times 64$&& 0.0030, 0.0040 &&   0.0220, 0.0270 && 0.2500, 0.3200   \\
\hline 
%&&&&&&&&& \\
%&&&&&&&&& \\
4.05      &&  $32^3 \times 64$&& 0.0030, 0.0060 &&   0.0150, 0.0180 && 0.2000, 0.2300     \\
          &&                  && 0.0080, 0.0120 &&   0.0220, 0.0260 && 0.2600, 0.3150     \\
\hline \hline
\end{tabular}
\end{center}
\caption{\it Lattice size, light ($=$~sea), strange- and charm-like bare quark mass values 
used in the analysis presented in this work. The number of correlator measurements was 240 
in all cases, but for $\beta=4.05$, where it was 130. The $r_0/a$-values $4.46(3)$, $5.22(2)$ and 
$6.61(3)$ are employed at $\beta=3.8$, $\beta=3.9$ and $\beta=4.05$, respectively~\cite{DFHUW_Latt07}. 
The overall scale and the light quark mass are set by the experimental values of $f_\pi$ and $m_\pi$
via chiral fits of the pseudoscalar meson mass and decay constant data in the light quark 
sector~\cite{Latt08,DFHUW_Latt07}. Here we use $\hat\mu_{u/d} = 3.6(3)$~MeV and $r_0=0.433(14)$~fm. 
} 
\label{tab:simpar}
\end{table}

In order to determine the coefficient functions $\eta_j$ we proceed as follows. Let us make for the smallest 
$\hat\mu_h$ value the choice $\hat\mu_h^{(1)}=1.230$~GeV (we recall, we are referring to the $\overline{MS}$ 
scheme at the scale of 2~GeV). Fixing $\lambda =1.278$ (see below) and, in this exploratory study, $N=4$, 
we shall successively consider the $h$-quark masses  
\beqn
&&\hat\mu_h^{(1)}= 1.230~{\mbox{GeV}}\, , \nn\\
&&\hat\mu_h^{(2)}= \lambda \hat\mu_h^{(1)} = 1.572~{\mbox{GeV}}\, , \nn\\
&&\hat\mu_h^{(3)}= \lambda^2 \hat\mu_h^{(1)} = 2.009~{\mbox{GeV}}\, ,\nn\\
&&\hat\mu_h^{(4)}= \lambda^3 \hat\mu_h^{(1)} = 2.568~{\mbox{GeV}}\, .
\label{MUV}\eeqn
Actually at each lattice spacing we will be dealing with the dimensionless quantities
$\hat\mu_h^{(j)}r_0$, $j=1,2,3,4$, of which the numbers quoted in eq.~(\ref{MUV}) 
represent the central values in physical units. Uncertainties on $r_0/a$ and $Z_P^{-1}$ (present 
at the level of about 3\%) will be taken into account in the final error analysis. 

\begin{figure}[!hbt]
\centerline{\includegraphics[scale=0.50,angle=-90]{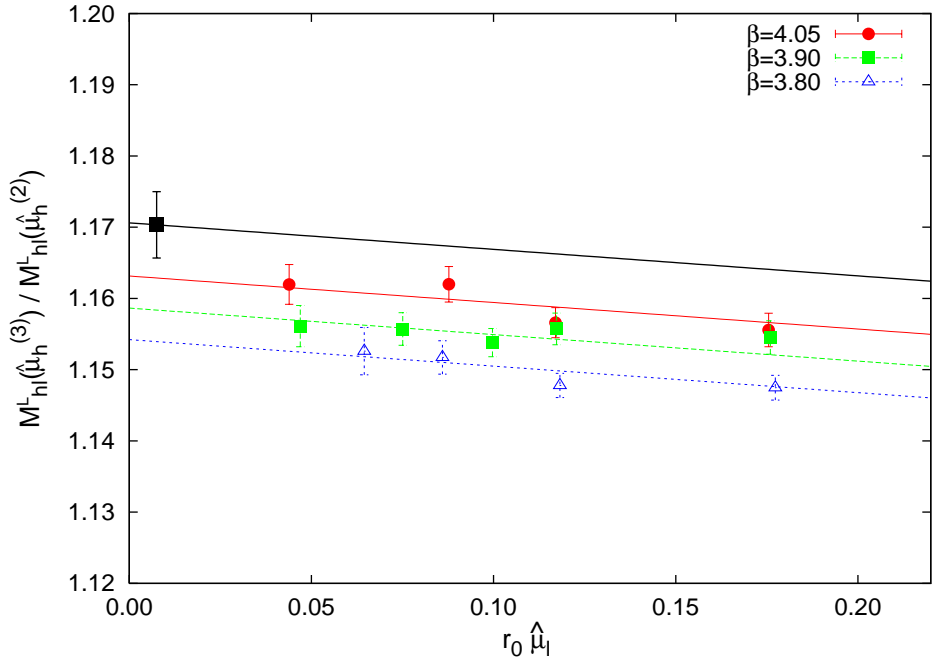}}
\caption{\it Lattice spacing and $\hat\mu_\ell$ dependence of the ratio 
$M_{h\ell}^L(\hat\mu_h^{(3)};\hat\mu_\ell,a)/M_{h\ell}^L(\hat\mu_h^{(2)};\hat\mu_\ell,a)$.
The black square with its error is the combined continuum and chirally ($\hat\mu_\ell \to \hat{\mu}_{u/d}$) 
extrapolated value. Here and in all the following figures uncertainties possibly affecting the value of the variable 
in the horizontal axis are propagated to the quantity plotted on the vertical axis.}
\label{fig:Mrat3to2}
\end{figure}

{}From the set of the ETMC simulation data~\cite{ETMCD} with parameters detailed in Table~\ref{tab:simpar}, 
we extract the values of the $h\ell$ pseudoscalar meson masses that correspond to the $\hat\mu_h$ values 
listed in~(\ref{MUV}). With these masses we construct the lattice ratios~(\ref{RATM}) on which a combined 
continuum and chiral fit is performed. As we hoped, ratios appear to have a mild dependence on the light quark 
mass $\hat{\mu}_\ell$ and small cutoff effects, as seen for instance in fig.~\ref{fig:Mrat3to2}. This makes our 
continuum and chiral fit straightforward and numerically robust. 
 
\begin{figure}[!hbt]
\centerline{\includegraphics[scale=1.0]{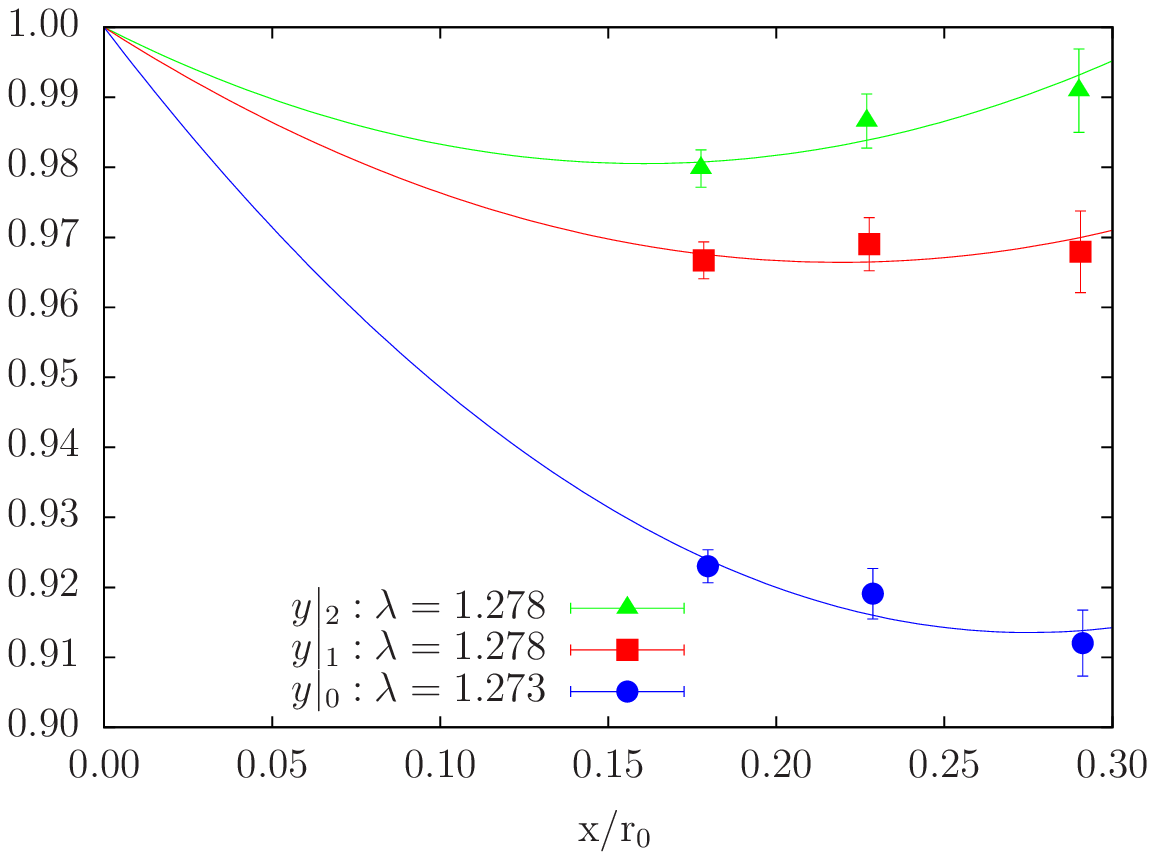}} 
\caption{\it Continuum data for $y|_0$ (blue dots), $y|_1$ (red squares), $y|_2$ (green triangles).  
The corresponding best fit curves are drawn with $\lambda=1.273$ (lower curve, in blue) and $\lambda=1.278$ 
(middle curve, in red and upper curve, in green). In all cases $\mu_\ell \to \mu_{u/d}$.}
\label{fig:fig1}
\end{figure}

The red squares in fig.~\ref{fig:fig1} represent the numbers $y^{(n)}_1=y(x^{(n)},1.278;\hat\mu_{u/d})|_1$, 
$n=2,3,4$, computed at the $x^{(n)}=1/\hat\mu_h^{(n)}$ values in the list~(\ref{MUV}) with the ratios 
${\rho(\log \hat\mu_h^{(n-1)})}/{\rho(\log \hat\mu_h^{(n)})}$ computed at LL order (i.e.\ $p=1$ in 
eq.~(\ref{LIM})). The best fit through the red squares and the point at $x=0$ determines the values 
of the $\eta_j$ coefficients ($j=1,2$) and yields the middle (red) curve in the figure. 
We note that a second order polynomial in $x$ is necessary to get a good fit to the data 
(a straight line forced to pass through the point $y=1$ at $x=0$ would have a very large $\chi^2$). 
The quadratic fit gives for the quantities $r_0\eta_{1}$ and $r_0^2\eta_{2}$ numbers of order unity, 
in agreement with the standard assumptions underlying HQET. 

\begin{figure}[!hbt]
\centerline{\includegraphics[scale=0.50,angle=-90]{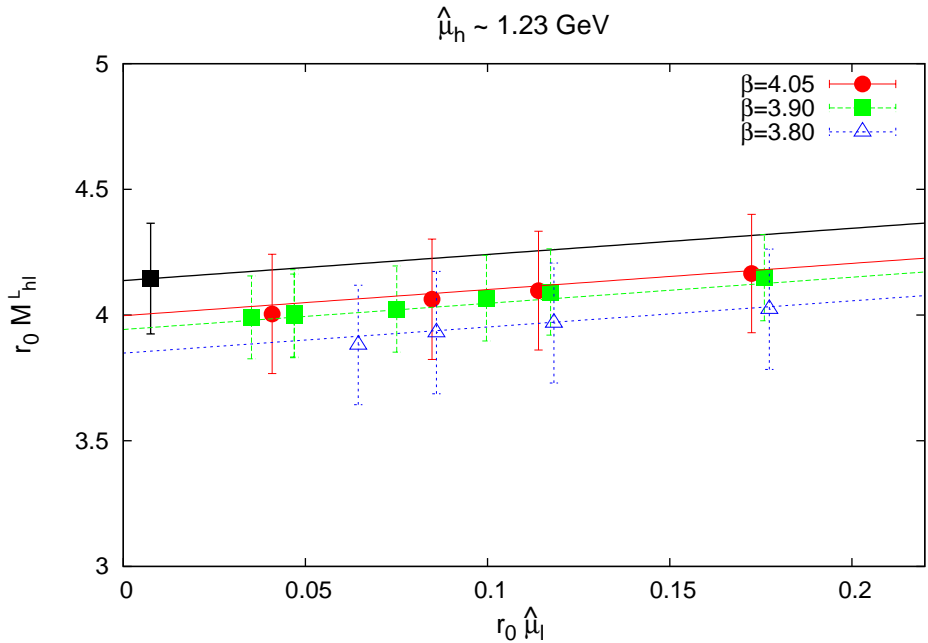}}
\caption{\it The $M_{h\ell}^L(\hat\mu_h^{(1)};\hat\mu_\ell,a)$ lattice data extracted from the 
simulations detailed in Table~\ref{tab:simpar}. The black square with its error is the continuum and 
chirally extrapolated value giving $M_{hu/d}(\hat\mu_h^{(1)})=1.89(10)$~GeV.} 
\label{fig:fitm}
\end{figure}

\begin{figure}[!hbt]
\centerline{\includegraphics[scale=1.0]{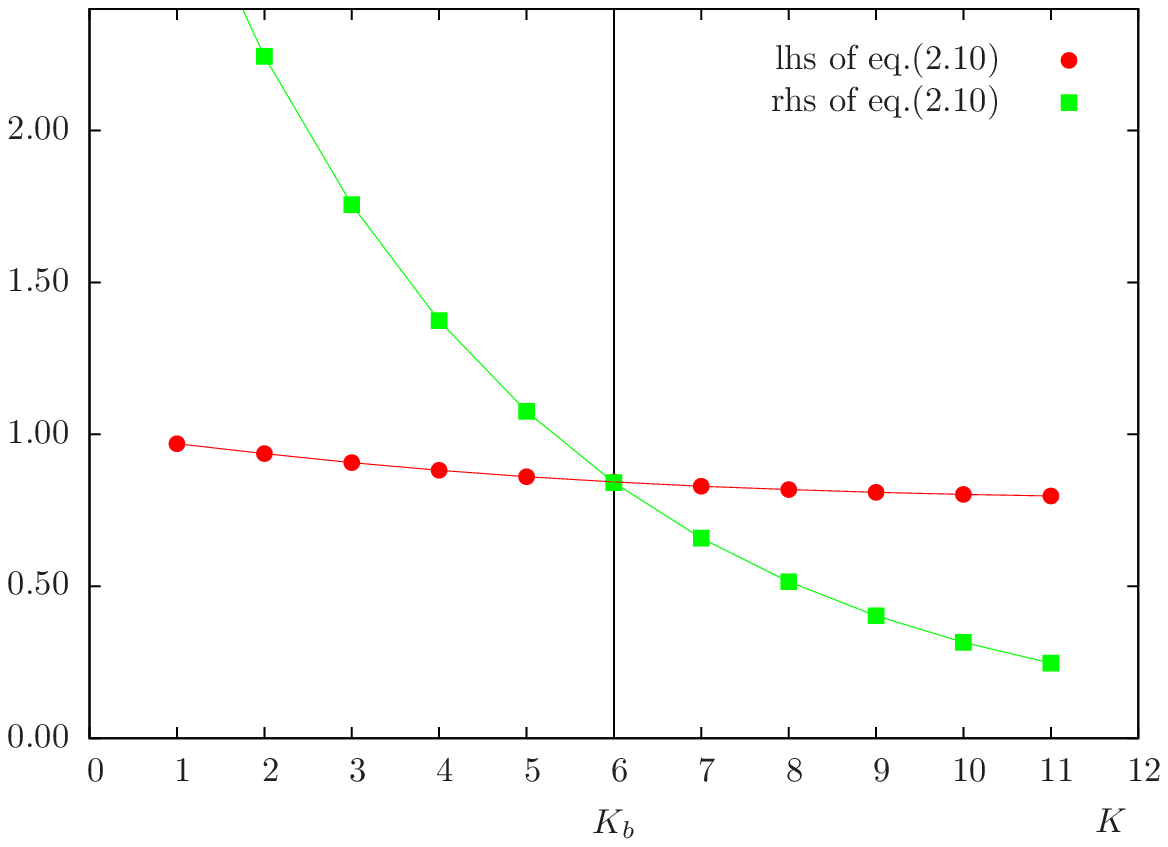}}
\caption{\it The numerical solution of eq.~(\ref{ITER}) giving $K_b= 6$.}
\label{fig:fig2}
\end{figure}

At this stage, having in our hands the quantities $y^{(n)}_1=y(x^{(n)},1.278;\hat\mu_{u/d})|_1$ for any $n$ 
(actually for any $x$), the iterative formula 
\beq
y_1^{(2)} y_1^{(3)}\cdots y_1^{(K+1)}=\lambda^{-K} 
\frac{M_{hu/d}(\hat\mu_h^{(K+1)})}{M_{hu/d}(\hat\mu_h^{(1)})} \cdot
\Big{[}\frac{\rho(\log \hat\mu_h^{(1)})}{\rho(\log \hat\mu_h^{(K+1)})}\Big{]}_{p=1}\, , 
\label{ITER}
\eeq
should be looked at as a relation between the mass of the $hu/d$-meson, $M_{hu/d}(\hat\mu_h^{(K+1)})$, 
and the corresponding heavy quark mass $\hat\mu_h^{(K+1)}$, which is fully explicit if the initial, 
triggering value $M_{hu/d}(\hat\mu_h^{(1)})$ is assigned. The latter can be accurately measured, as 
$\hat\mu_h^{(1)}$ lies in the well accessible charm quark mass region. We show in fig.~\ref{fig:fitm} 
the quality of the continuum and chiral extrapolation of the triggering mass lattice data. Once this 
number is known, determination of the $b$-quark mass is tantamount to find the value of $K$ at which 
$M_{hu/d}(\hat\mu_h^{(K+1)})$ takes the experimental $B$-meson mass value, $M_B$. Calling 
$K_b$ the solution of the resulting eq.~(\ref{ITER}) (as shown in fig.~\ref{fig:fig2}, we find $K_b=6$), 
one gets for $\hat\mu_b$ the simple formula (valid for renormalized as well as bare masses) 
\beq
\hat\mu_b=\lambda^{K_b}\hat\mu_h^{(1)}\, .
\label{MB}
\eeq 

A few related remarks are important here. 1) It is not really necessary to have the lattice 
$h\ell$ pseudoscalar meson masses computed at values of $\hat\mu_h$ matched exactly as indicated 
in eq.~(\ref{MUV}). A $\mu_h$ interpolation between nearby $M_{h\ell}^L$ masses can be carried out if 
necessary. This is what we have actually done in the numerical study we present in this 
paper. 2) It is not a priori guaranteed that eq.~(\ref{ITER}) can be solved for an integer value of the 
exponent $K$. This is not a problem, however, as one can always retune the parameter $\lambda$ 
(and at the same time readjust the values in the sequence~(\ref{MUV})), so as to end up with an 
integer for $K_b$ (this is the reason why the peculiar value $\lambda=1.278$ was chosen). 
Alternatively one could adjust the starting value of the heavy quark mass or both. 
3) A detailed discussion of the numerical analysis will be given in a forthcoming
publication~\cite{FUT}. Here we only mention that
a simple SU($N_f=2$) chiral perturbation theory NLO-formula was used to model the $\hat\mu_\ell$
dependence of the triggering $h\ell$ meson mass and $y$-ratios, while O($a^2$) effects have been
parameterized (at each $\hat\mu_h$) by $\hat\mu_\ell$ independent terms. A few further details
on this point are given in Appendix~A.

Following the procedure outlined above, one finds the result given in eq.~(\ref{UQRES1}), or 
equivalently the renormalization group invariant (RGI) value 
\beq 
\hat\mu_b^{RGI,N_f=2} = 7.6(5)~{\mbox{GeV}}\, , 
\label{MUBARE}
\eeq 
where in the running only two flavours are assumed to be active and the conventions 
of ref.~\cite{GL} for the RGI quark mass have been used. 

\subsection{Discussion and error budget}
\label{sec:TLNLL}

It is important to check the degree of reliability of the key smoothness assumption we have been implicitly
making on the function $y(x)$ and test the sensitivity of the procedure and its result~(\ref{MB}) to the
order of PT at which the expansion of $\rho$ is truncated. To this end we have repeated the entire analysis 
above using for $\rho$ both a lower (tree-level) and a higher (NLL) order perturbative approximation 
in place of the previously employed LL order truncation. We recall that in the large $h$-quark mass 
limit $y|_{0}=y(x,\lambda;\hat\mu_{u/d})|_{0}$ and $y|_{2}=y(x,\lambda;\hat\mu_{u/d})|_{2}$ 
approach 1 (see eq.~(\ref{LIM})) with corrections O($1/\log x$) and O($1/(\log x)^3$), respectively.

One finds a (very) little shift in the value of $\lambda$ necessary for the solution of the corresponding iterative 
equation~(\ref{ITER}) to be an integer (from 1.278 to 1.273) if we go from $y|_1$ to $y|_0$. The shift is 
instead totally negligible within our statistical accuracy if we move from $y|_1$ to $y|_2$. 

The fits to $y|_0$ and $y|_2$ data are shown in fig.~\ref{fig:fig1} (lower blue and upper green curve) together 
with the fit to $y|_1$ data (middle red curve). One clearly sees that the $y|_p$ curves tend to become flatter and 
flatter as we move from $p=0$ to $p=2$. As for the values of the $b$-quark mass, instead, the number extracted 
from $y|_0$-ratios is only about 2\% smaller than the one obtained using the $y|_1$-ratios. The difference between 
the latter and the one extracted from $y|_2$-ratios is smaller than 1\%. 

The stability of the value of the $b$-quark mass with varying $p$ should not come as a surprise. 
It is enough to notice that, if the $y|_p$-ratios were exactly known, for a generic value of 
$p$ one would get (recall eq.~(\ref{ITER})) 
\beqn
&&\Big{[}\frac{\rho(\log \hat\mu_h^{(K+1)})}{\rho(\log \hat\mu_h^{(1)})}\Big{]}_{p} \cdot
y_p^{(2)} y_p^{(3)}\cdots y_p^{(K+1)}=\nn\\
&&=y_0^{(2)} y_0^{(3)}\cdots y_0^{(K+1)}=\lambda^{-K} 
\frac{M_{hu/d}(\hat\mu_h^{(K+1)})}{M_{hu/d}(\hat\mu_h^{(1)})} \, ,
\label{ITER1}
\eeqn
as all the intermediate $\rho$ factors (except the first and the last) cancel out in the l.h.s.\ leaving 
behind simply the product of the $y|_0$-ratios. 
The small $p$ dependence we have found in the value of $\lambda$ 
(hence in $\hat\mu_b$) is due to the slightly different level of accuracy
by which the $y|_p$-ratios (which instead significantly depend on $p$) 
can be described by a polynomial in $x$ with the lowest
order coefficient set to unity. In this respect, increasing $p$
is expected to improve the quality of the ansatz~(\ref{ANLYTICY}) and reduce 
the systematic error associated to it. This is so until $p$ becomes so ``large'' 
that the accuracy of the $\rho$ estimate gets spoiled by
the renormalon ambiguity in its perturbative expansion~\cite{MarSac96}.

To account for the truncation to LL of the perturbative expansion for $\rho$ we have 
conservatively decided to attribute to the $b$-quark mass value a systematic error of 1\%, 
which is added in quadrature to the other errors discussed below, leading to the total error quoted 
in eq.~(\ref{UQRES1}) (and~(\ref{MUBARE})). 

\subsubsection{Error budget}
\label{sec:EB}

The total error we attribute to the $b$-quark mass results~(\ref{UQRES1}) and~(\ref{MUBARE}) 
takes into account a number of statistical and systematic effects which we now briefly illustrate.  
The relative error on the product of the continuum $y$-ratios in the l.h.s.\ of eq.~(\ref{ITER}) is only about 1\%, 
whereas the pseudoscalar meson mass in the charm region ($M_{hu/d}(\hat\mu_h^{(1)})$ in the r.h.s.\ of 
eq.~(\ref{ITER})) contributes a relative error of about 5\%. These errors are the result of our statistically 
limited knowledge of $h \ell$-meson correlators, $r_0/a$ and $Z_P$~\footnote{At the moment the statistical 
error on $Z_P$ quoted by ETMC is about $3\%$ at the simulated lattice spacings.} as well as of a number of further 
systematic errors. Among the latter we mention those coming from the fit ansatz underlying the combined 
continuum and chiral ($\hat\mu_\ell \to \hat\mu_{u/d}$) extrapolation, the error due to the $x$-interpolation 
to the $b$-mass point, as well as the (tiny) error inherent the numerical solution of eq.~(\ref{ITER}) 
(giving $K_b=6$ and $\lambda=1.278$). As we discussed above, the effect on $\hat\mu_b$ due to the truncation 
of the $\rho$ perturbative series to order $p$ is very small, not larger than 1\%. 
Another .5-1\% systematic error comes from the possible (neglected) 
logarithmic dependence of the $\eta_j$, $j=1,2$ coefficients. 
The relative uncertainty on $\eta_j$ associated with these effects can be estimated to be 
O($\alpha_s(1/x))\sim 10-15$\%, a number which is never larger 
than the statistical errors on their best fit values.
%%%
Finally cross-correlations between the different quantities (stemming from common 
ensembles of gauge configurations) are as usual taken into account by a bootstrap error 
analysis. Further technical aspects of the error analysis are deferred to ref.~\cite{FUT}. 

The information provided in figs.~\ref{fig:Mrat3to2}, \ref{fig:fig1} and~\ref{fig:fitm} about the $a^2$, 
$\hat\mu_\ell$ and $x$ dependence of the intermediate quantities entering our analysis as well 
as about the precision in solving eq.~(\ref{ITER}) (see fig.~\ref{fig:fig2}) shows that the global 
systematic uncertainty is well within (or below) our present statistical errors. 

We conclude by observing that, as expected, our results for the $b$-quark mass 
(and $f_B$ or $f_{B_s}$ discussed in the next section) do not significantly depend on the
value of the intermediate quantity $r_0$ which is only employed to ease continuum 
extrapolations, while the physical scale is ultimately set by $f_\pi$. 

\subsection{$b$ and $c$ quark masses}
\label{sec:BCQM}

Although not necessary, the phenomenological value of the $D$-meson mass could have been  
used as a triggering mass. In this case $\mu_h^{(1)}$ would have to be identified with $\mu_c$. 

We note in this context that, since we get (see the black square in fig.~\ref{fig:fitm}  
$M_{hu/d}(\hat\mu_h^{(1)})=1.89(10)$~GeV, i.e.\ a number that practically coincides with 
the experimental value of $M_D$ ($M_{D^0}=1.865$~GeV~\cite{REVSM}), 
our method immediately yields for the charm mass the estimate $\hat\mu_c=1.23(06)$~GeV.
For the phenomenologically important $b$- over $c$-mass ratio we then get 
\beq
\frac{\hat \mu_b}{\hat \mu_c}=4.31(24)\, ,
\label{BOCM} 
\eeq
in very good agreement with other estimates~\footnote{It is interesting to compare our unquenched result for the ratio of RGI 
masses $m_c^{RGI,N_f=2}/m_b^{RGI,N_f=2}=0.232(13)$ (the inverse of eq.~(\ref{BOCM})) with the corresponding quantity  
$[m_c^{RGI,N_f=0}=1.654(45)~{\rm GeV}]/[m_b^{RGI,N_f=0}=6.758(86)~{\rm GeV}]=0.245(7)$ 
determined using the quenched data of ref.~\cite{RS} (for the $c$-mass) and~\cite{DELM} (for the $b$-mass). 
More recently the work of ref.~\cite{CHENEW} has appeared where the number 
$m_c^{\overline{MS},N_f=4}(3~{\rm GeV})/m_b^{\overline{MS},N_f=4}(10~{\rm GeV})=0.273(3)$ 
is quoted. This value is well consistent with our result~(\ref{BOCM}) which translates into 
$m_c^{\overline{MS},N_f=2}(3~{\rm GeV})/m_b^{\overline{MS},N_f=2}(10~{\rm GeV})=0.274(15)$.}. 

In closing this section we note that an independent determination of the $b$-quark mass can be 
obtained repeating the same analysis as before but using $M_{B_s}$, instead of $M_B$, and replacing 
$\hat\mu_{u/d}$ with $\hat\mu_s$. By doing that we find a result which is fully consistent with the 
one in eq.~(\ref{MUBARE}). Alternatively, and perhaps more interestingly, one could use $K_b$ as 
determined from $M_B$ to predict $M_{B_s}$, or better the ratio $M_{B_s}/M_B$, by the method we 
are proposing in this paper. Such an analysis is in progress and will be presented elsewhere~\cite{FUT}. 

\section{$f_B$ and $f_{B_s}$ decay constants}
\label{sec:METHF}

A strategy very similar to the one outlined in sect.~\ref{sec:METHB} can be employed to extract 
accurate values of the $f_B$ and $f_{B_s}$ decay constants from available lattice data. In analogy 
with what we have done before, one should now take 
\beq
z(x,\lambda;\hat\mu_\ell)=\lambda^{1/2} \frac{f_{h\ell}(1/x)}{f_{h\ell}(1/x\lambda)}
\cdot\frac{C^{stat}_A(\log(x\lambda))}{C^{stat}_A(\log x)} 
\frac{[\rho(\log x)]^{1/2}}{[\rho(\log \lambda x)]^{1/2}}\label{RATMZ}
\eeq
with the (continuum limit) shorthand notation
\[
f_{h\ell}(1/x)\equiv f_{h\ell}(1/x,\hat\mu_\ell)\, , 
\]
where the quark mass $\hat\mu_\ell$ must be extrapolated to either $\hat\mu_{u/d}$ or to the appropriate 
strange quark mass value, $\hat\mu_s$, depending on whether one wants to compute $f_B$ or $f_{B_s}$. 

\begin{figure}[!hbt]
\centerline{\includegraphics[scale=0.50,angle=-90]{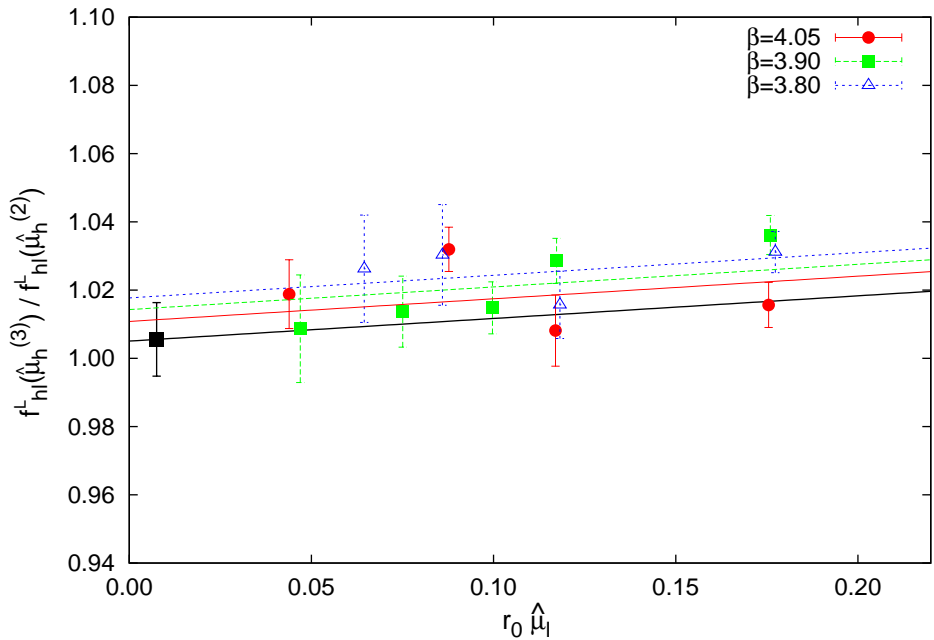}}
\caption{\it Lattice spacing and $\hat\mu_\ell$ dependence of the ratio 
$f_{h\ell}^L(\hat\mu_h^{(3)};\hat\mu_\ell,a)/f^L_{h\ell}(\hat\mu_h^{(2)};\hat\mu_\ell,a)$.
The black square with its error is the combined continuum and chirally ($\hat\mu_\ell \to \hat{\mu}_{u/d}$) 
extrapolated value.}
\label{fig:Frat3to2}
\end{figure}

The form of the function $z(x,\lambda;\hat\mu_\ell)$ is dictated by the continuum asymptotic formula 
\beq
\lim_{x\to 0} \sqrt{\frac{\rho(\log x)}{x}}\frac{f_{h\ell}(1/x)}{C^{stat}_A(\log{x})} 
= {\mbox{constant}} \neq 0\, ,
\label{HQM}
\eeq
which follows by matching HQET to QCD~\cite{HQET,HQET1}.  The presence of the factor $C^{stat}_A$  
comes from the fact that in HQET the axial (and vector) current needs to be renormalized.
The renormalization constant $C^{stat}_A$ is known in PT up to three loops~\cite{ZAPT}. The ratio of 
$\rho$ factors (raised to the appropriate power) is there to convert $\overline{MS}$ heavy quark masses 
to pole masses (see eq.~(\ref{PMA})). 

As before, the function~(\ref{RATMZ}) has been defined so as to fulfill the exact asymptotic constraint 
\beq
\lim_{x\to 0} z(x,\lambda;\hat\mu_{\ell})=1\, , \label{RATLEZ}
\eeq
from which the small $x$ expansion 
\beq
z(x,\lambda,\hat\mu_{\ell})=1+\zeta_1(\log x,\lambda;\hat\mu_{\ell})x
+\zeta_2(\log x,\lambda;\hat\mu_{\ell}) x^2 \, , \label{ANLYTICF}
\eeq
follows. Again the coefficients $\zeta_j(\log x,\lambda;\hat\mu_{\ell})$, $j=1,2$, are 
smooth functions of $\log x$ which tend to zero as $\lambda \to 1$. 

In analogy with what we did in sect.~\ref{sec:METHB} in determining the $b$-quark mass, 
with the purpose of  
checking the robustness of the procedure, we shall take $C_A^{stat}$ and $\rho$ at increasing orders in PT, 
from tree-level up to NLL order, and construct $z|_p$-ratios endowed with the asymptotic behaviour 
\beq 
z(x,\lambda;\hat\mu_{u/d})\Big{|}_{p}-1
~\stackrel{x \to 0}{\sim}~ {\rm O}\Big{(}\frac{1}{(\log x)^{p+1}}\Big{)}\, , \label{ZLIM}
\eeq
Just like in the case of the determination of the $b$-quark mass, the values of $f_B$ or $f_{B_s}$ that 
we shall extract will be almost independent of the PT truncation order. 

\begin{figure}[!hbt]
\centerline{\includegraphics[scale=1.0]{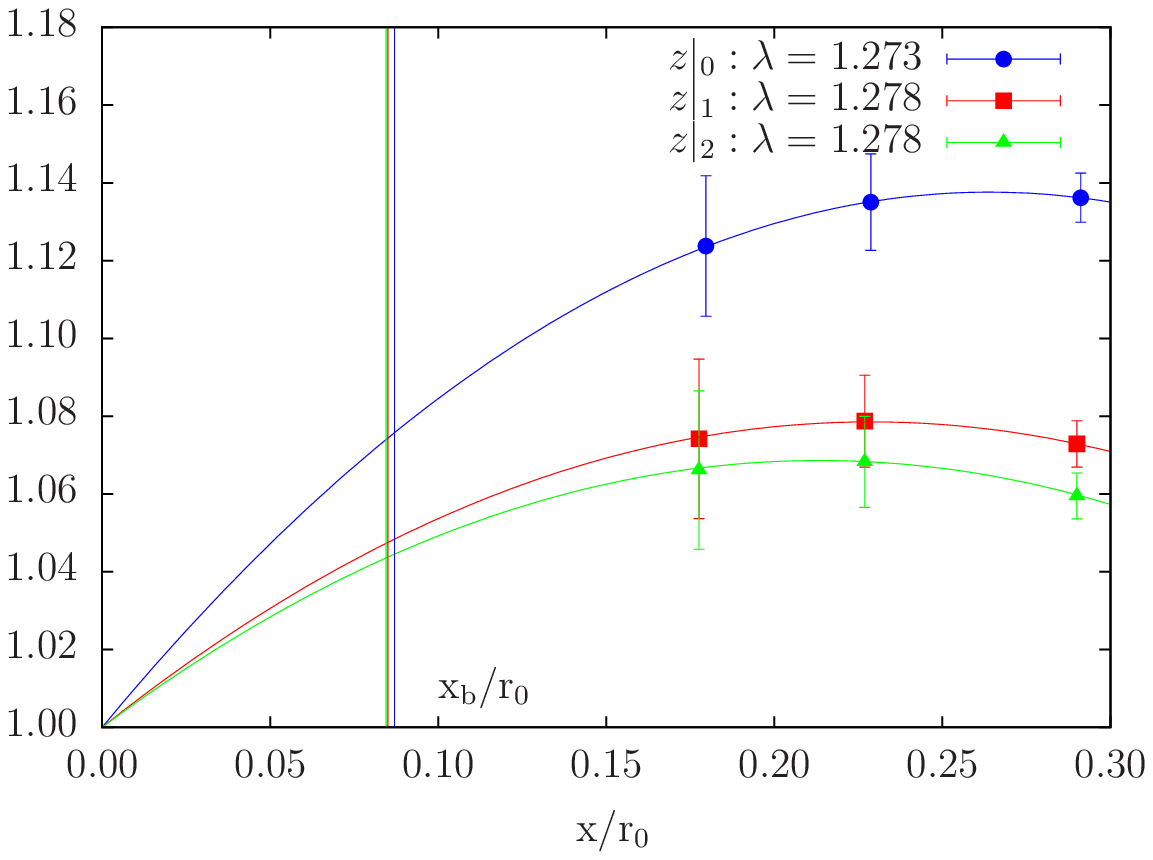}}
\caption{\it Continuum data for $z|_0$ (blue dots), $z|_1$ (red squares), $z|_2$ (green triangles). 
The corresponding best fit curves are drawn with $\lambda=1.273$ (upper curve, in blue) and $\lambda=1.278$ 
(middle curve, in red and lower curve, in green). In all cases $\mu_\ell \to \mu_{u/d}$. 
The blue and red vertical lines represent the location of the $b$-mass as extracted from $y|_0$ 
and $y|_1$ data (with $\lambda=1.273$ and $\lambda=1.278$), respectively. The green vertical line is 
practically on top of the red line and it is not visible.}
\label{fig:fig3}
\end{figure}

\subsection{Implementing the method. The case of $f_B$}
\label{sec:IMFB}

The $z$-ratios~(\ref{RATMZ}) have been evaluated at the reference $h$-quark masses of the list~(\ref{MUV})
for each of the lattice spacings and light quark mass values given in Table~(\ref{tab:simpar}). When we 
perform the continuum and chiral extrapolation of the ETMC lattice data for the ratios~(\ref{RATMZ}) of 
$h\ell$ pseudoscalar meson decay constants (again based on simple chiral NLO-formulae 
supplemented 
with $\hat\mu_\ell$-independent O($a^2$) corrections -- see Appendix~A), 
as hoped, a rather smooth behaviour is found  
since most of the $a^2$ and $\hat\mu_\ell$ dependence gets canceled in taking the ratio. The observable 
dependence on $\hat\mu_\ell$ and $a^2$ is mild and/or hardly significant within our present statistical 
errors (see e.g.\ fig.~\ref{fig:Frat3to2}). 

\begin{figure}[!hbt]
\centerline{\includegraphics[scale=0.50,angle=-90]{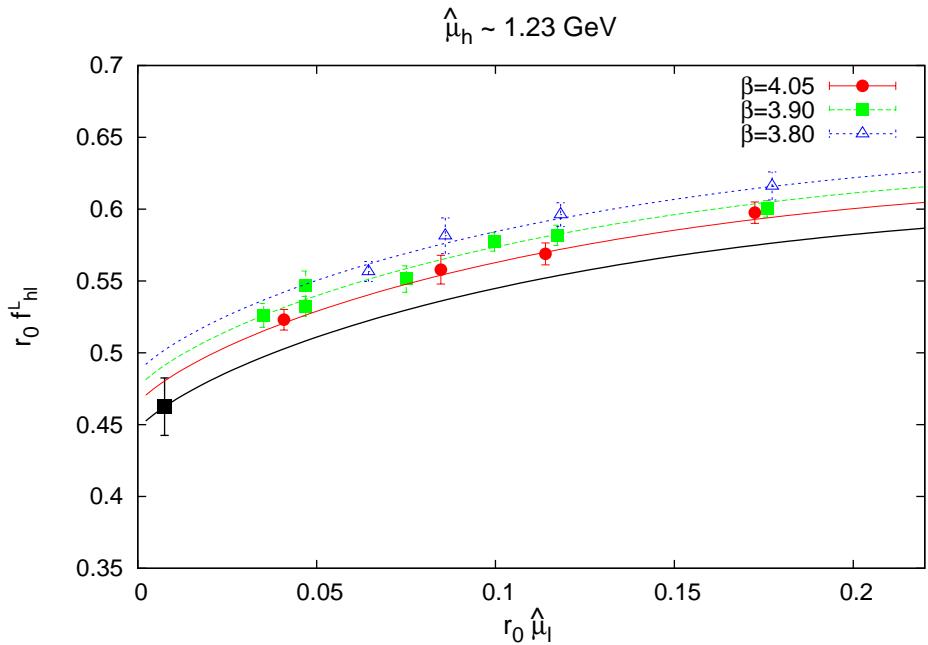}}
\caption{\it The $f_{h\ell}^L(\mu_h^{(1)};\hat\mu_\ell,a)$ lattice data extracted from the 
simulations detailed in Table~\ref{tab:simpar}. The black square 
with its error is the continuum and chirally extrapolated value giving 
$f_{hu/d}(\hat\mu_h^{(1)},\hat\mu_{u/d})= 211(9)$~MeV.}
\label{fig:fitf}
\end{figure}

{}From the structure of eq.~(\ref{RATMZ}) one derives the iterative formulae (analogous to eq.~(\ref{ITER})
with $z^{(n)}_p\equiv z(x^{(n)},\lambda;\hat\mu_{u/d})|_p$) 
\beqn 
&&z_p^{(2)} z_p^{(3)}\cdots z_p^{(K+1)}=\label{ITERZ}\\
&&=\lambda^{K/2} \frac{f_{h\ell}(\hat\mu_h^{(K+1)})}{f_{h\ell}(\hat\mu_h^{(1)})} \cdot
\Big{[}\frac{C^{stat}_A(\log \hat\mu_h^{(1)})}{C^{stat}_A(\log\hat\mu_h^{(K+1)})}
\sqrt{\frac{\rho(\log\hat\mu_h^{(K+1)})}{\rho{(\log \hat\mu_h^{(1)})}}}\Big{]}_{p}\, ,\quad p=0,1,2\, .\nn
\eeqn 
Similarly to what we did in fig.~\ref{fig:fig1}, we collect in fig.~\ref{fig:fig3} continuum and chirally 
extrapolated data for $z|_p$, $p=0,1,2$ and best fit curves through these data and the value at $x=0$.
Thus, for instance, the middle (red) curve is the parabola (eq.~(\ref{ANLYTICF})) which best 
fits the values of $z_1^{(n)}=z(x^{(n)},1.278;\hat\mu_{u/d})|_1$, $n=2,3,4$, at the heavy 
quark masses~(\ref{MUV}). The red vertical line marks the position $x_b$ which corresponds to the 
previously determined value of $\hat\mu_b$ (eq.~(\ref{MUBARE})) and crosses the curve at the 
point $z_1^{(K_b)}=z(x_b,1.278;\hat\mu_{u/d})|_1$. With the help of this number and the values of 
$z_1^{(j)} $ for $4<j\leq K_b+1$, eq.~(\ref{ITERZ}) provides a determination of $f_{hu/d}(\hat\mu_b)$ 
in terms of $f_{hu/d}(\hat\mu_h^{(1)})$ (with LL-accurate fit for the $z$-ratios). 
As observed before, the latter does not necessarily has to be identified with the phenomenological 
value of $f_D$ for the method to work, as what we actually need to know is the dependence 
of $f_{hu/d}(\hat\mu_h)$ on $\hat\mu_h$ at around the charm mass. Nevertheless, since, as remarked 
in sect.~\ref{sec:TLNLL}, $M_{hu/d}(\hat\mu_h^{(1)})$ coincides with the experimental value of $M_D$, 
we are in position of evaluating $f_D$, obtaining $f_D=f_{hu/d}(\hat\mu_c)= 211(9)$~MeV, 
compatible with the result $f_D= 197(9)$~MeV given in ref.~\cite{VCS}. 
The latter was obtained in the standard way (see ref.~\cite{ETMCD}) from
the same ETMC gauge configuration ensembles, but with a rather different analysis
method where the meson masses rather than the renormalized quark masses were
kept fixed as $a \to 0$, resulting in somewhat different statistical (no use of $Z_P$)
and systematic errors as compared to the present study.
For the present computation of $f_{hu/d}(\hat\mu_h^{(1)})= f_D$ the
quality of the continuum and chiral extrapolation of our lattice data
is shown in fig.~\ref{fig:fitf}. Taking as triggering value the continuum and chirally extrapolated 
value of the pseudoscalar decay constant computed at $\hat\mu_h^{(1)}$, we get
\beq
f_B=f_{hu/d}(\hat\mu_b)=194(16){\mbox{MeV}} \, ,
\label{FB}\eeq
which is precisely the result~(\ref{UQRES2}). 

\subsection{Discussion and $f_B$ error budget}
\label{sec:EBF}

To test the reliability of the interpolation of our trial functions, $z(x)$, to $b$-mass point,
we have explicitly checked the stability of the result~(\ref{FB}) to the order of PT at which 
the expansion of $\rho^{{1/2}}/C_A^{stat}$ is truncated. For this purpose we have repeated 
the whole previous analysis employing values of $\rho^{{1/2}}/C_A^{stat}$ computed at tree-level 
($p=0$) and NLL ($p=2$) order. Upon comparing with the decay constant values obtained in these 
other ways we see that numbers obtained using the $z|_0$-ratios (upper blue curve 
in fig.~\ref{fig:fig3}) differ by less than 1\% from the value one gets from the red data (LL $z|_1$-ratios). 
If one employs the lower green data (coming from the NLL $z|_2$-ratios) the difference with the previous 
determination is totally negligible (about 0.1\%). This specific systematic effect on $f_B$ was hence 
conservatively estimated to be $\sim 0.5\%$ and added quadratically to the full error.

As in the case of the determination of the $b$-quark mass, the remarkable numerical stability of $f_B$ 
with varying $p$ can be traced back to the good quality of the interpolation ansatz~(\ref{ANLYTICF})
and the relation (again valid for exactly known $z|_p$-ratios) 
\beqn 
&&\Big{[}\frac{C^{stat}_A(\log\hat\mu_h^{(K+1)})}{C^{stat}_A(\log \hat\mu_h^{(1)})}
\sqrt{\frac{\rho{(\log \hat\mu_h^{(1)})}}{\rho(\log\hat\mu_h^{(K+1)})}}\Big{]}_{p} \cdot
z_p^{(2)} z_p^{(3)}\cdots z_p^{(K+1)}=\nn\\
&&=z_0^{(2)} z_0^{(3)}\cdots z_0^{(K+1)}=
\lambda^{K/2} \frac{f_{h\ell}(\hat\mu_h^{(K+1)})}{f_{h\ell}(\hat\mu_h^{(1)})}\, .\label{ITERZ1}
\eeqn 

\subsubsection{The $f_B$ error budget}
\label{sec:DEFB}

The total error we attribute to $f_B$ in eqs~(\ref{UQRES2}) and~(\ref{FB}) comes in almost equal
parts from the product of $z$-ratios in the l.h.s.\ of eq.~(\ref{ITERZ}) and the value of
$f_{hu/d}(\hat\mu_h^{(1)})$ and is a combination of statistical and systematic errors stemming
from the same sources already illustrated in the case of the $b$-quark mass in sect.~(\ref{sec:TLNLL}). As we saw 
above, the systematic error stemming from the truncation of the PT expansion of $\rho^{{1/2}}/C_A^{stat}$ 
has a negligible impact on $f_B$. 
Another 1-2\% systematic uncertainty comes from the possible (neglected) 
logarithmic dependence of the $\zeta_j$, $j=1,2$ coefficients. 
The relative uncertainty on $\zeta_j$ associated with these effects can be estimated to be 
O$(\alpha_S(1/x))\sim 10-15$\%, a number which, 
as in the case of the $\eta_j$'s entering our analysis for $\mu_b$,
is never larger than the statistical errors on their best fit values.
In any case inspection of figs.~\ref{fig:Frat3to2} to~\ref{fig:fitf} 
shows that all systematic uncertainties are smaller than our current statistical errors. 

\subsection{The case of $f_{B_s}$}
\label{sec:CFBS}

\begin{figure}[!hbt]
\centerline{\includegraphics[scale=1.0]{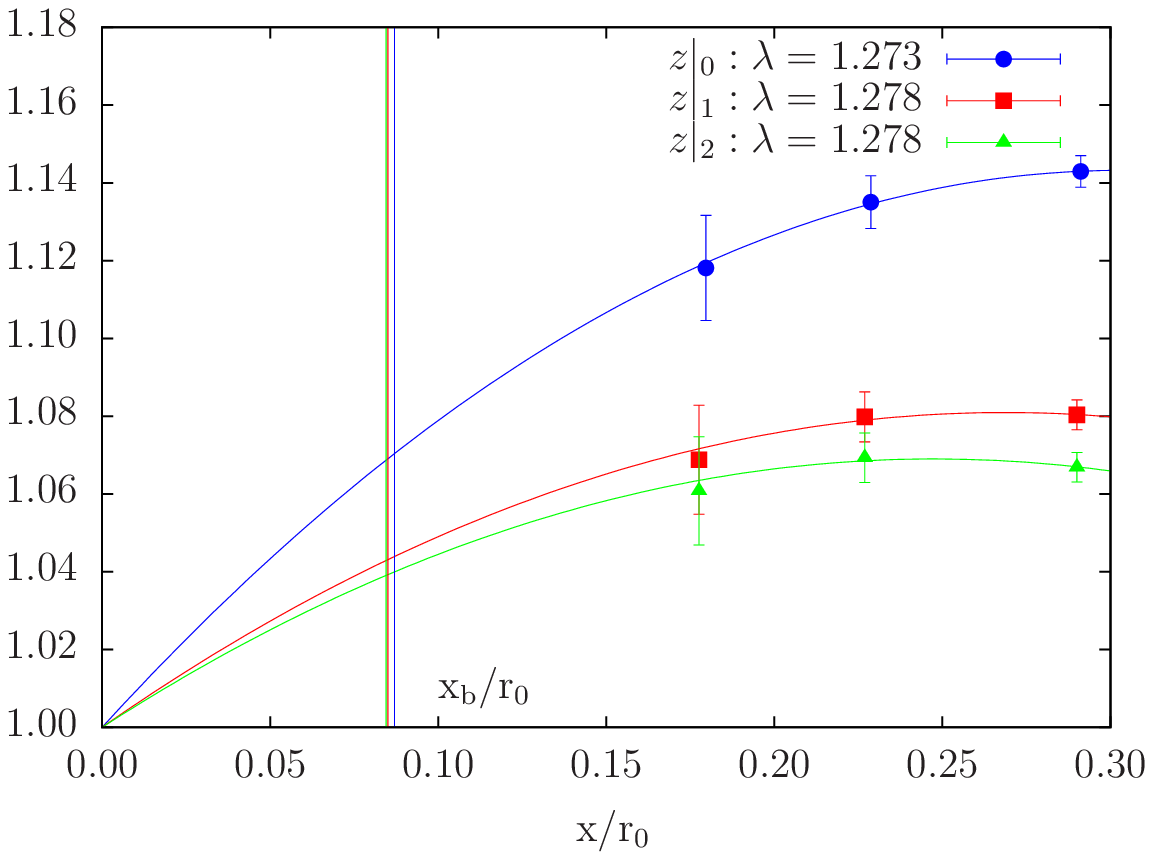}}
\caption{\it Same as fig.~\ref{fig:fig3} for $\mu_\ell \to \mu_s$.} 
\label{fig:fig4}
\end{figure}

In order to come up with a determination of $f_{B_s}$ one has simply to repeat the whole procedure setting 
$\mu_\ell\to\mu_s$. With reference to the value of $\hat\mu_s^{\overline{MS},N_f=2}(2~{\rm GeV})=99(7)$~MeV 
given in~\cite{ETMC_m_strange}, one finds 
(thanks to the equality of $\hat\mu_h^{(1)}$ with the charm quark mass) 
\beq
f_{D_s} = f_{hs}(\hat\mu_h^{(1)}) = 252(7)~{\rm MeV}
\label{fDs}
\eeq
and the best fit $z$-ratio curves shown in fig.~\ref{fig:fig4}. We may quote 
as our final result (see also eq.~(\ref{UQRES2}))
\beq 
f_{B_s}=f_{hs}(\hat\mu_b)=235(12)~{\mbox{MeV}}\, ,
\label{FBS}\eeq 
where errors are estimated as in the case of $f_B$. We also note that the result (\ref{fDs}) for
the decay constant of the $D_s$ meson is in agreement with that of ref.~\cite{VCS} and
contributes to further reduce the possible ``tension'' with the recent Cleo data reanalysis~\cite{CLEO_jan09}.

A more complete analysis of the many possibilities and refinements one can envisage will be presented in~\cite{FUT}.

As is clear by comparing the results for $f_B$ and $f_{B_s}$ (eqs.~(\ref{FB}) and~(\ref{FBS}), 
respectively), our method yields a significantly smaller error for the decay constant of the strange 
$B$-meson. The reason is that no large statistical fluctuations from the light ($u/d$) quark 
propagators nor (valence) chiral extrapolation uncertainties enter the computation of $f_{B_s}$. 
In view of this observation we remark that, if one would know with high accuracy 
the ratio $f_B/f_{B_s}$, a more precise determination of $f_B$ could be obtained by 
simply multiplying this number by $f_{B_s}$~\footnote{This observation is far from
original, see e.g.\ ref.~\cite{BECIR98}.}. Actually the quantity $f_B/f_{B_s}$ 
can be accurately computed by a simple generalization of the method discussed in this paper. 
It is indeed sufficient to consider the double ratio 
\beq
\hspace{-.02cm}
w(x,\lambda;\hat\mu_{u/d},\hat\mu_s)= [f_{h/ud}/f_{hs}](1/x) [f_{hs}/f_{hu/d}](1/ x\lambda ) =
\frac{z(x,\lambda;\hat\mu_{u/d})}{z(x,\lambda;\hat\mu_s)}\!\!
\label{FBOFBS}
\eeq
and follow the procedure we described before starting from the triggering quantity 
$[f_{hu/d}/f_{hs}](\hat\mu_h^{(1)})$. This kind of analysis is under way and will be 
discussed elsewhere~\cite{FUT}.

\section{Conclusions and outlook}
\label{sec:CONC}

In this paper we have proposed a novel strategy to determine $B$-physics parameters from currently 
available Wilson fermion simulation data. As a first test of the method, we have computed in (the continuum 
limit of) QCD with $N_f=2$ light dynamical quarks the (renormalized) $b$-quark mass as well as the 
$B$-meson decay constants, $f_B$ and $f_{B_s}$, employing the gauge configurations  
recently produced by the ETM Collaboration with maximally twisted Wilson fermion action~\cite{ETMCD}.

The method provides rather accurate numbers with errors that are dominated by the 
uncertainties related to the limited statistical accuracy by which the (two-point) $h\ell$ pseudoscalar
meson correlators and the quark mass renormalization constant, $Z_P^{-1}$, are evaluated. 
A better assessment of the systematic errors due to the limited knowledge of logarithmic
corrections can only come from data taken at quark masses larger than the ones displayed in eq.~(\ref{MUV}).

In several respects the present feasibility study could benefit from the nice properties (particularly 
O($a$) improvement~\cite{FR1,FR2}) of maximally twisted Wilson fermions. Indeed, an important 
feature of the present computation is the pretty good control we have of cutoff effects, which 
(judging from the spread between values at the coarsest lattice spacing and those at the continuum limit) 
are always smaller than 10\%. This is so both for the triggering quantities at the charm mass scale 
and for the $y$- and $z$-ratios, which involve higher quark masses (up to twice the charm mass). 

It is also interesting to note that the whole procedure only relies on the use of physical 
quantities that can be easily determined from lattice simulations, while the need 
for a renormalization step is limited to establishing the relation between the 
renormalized charm-like mass and the values of the triggering pseudoscalar meson mass. 
Fixing this relation requires the knowledge of $Z_P$. No extra renormalization factor is 
needed for the calculation of the decay constants of interest if maximally twisted fermions 
are used as the charged axial currents are exactly conserved at finite lattice spacing.  

There is a lot of room for improvement in the application of the method, like reducing the 
statistical error of the correlation functions, using several, suitably smeared meson sources, 
increasing the accuracy by which $Z_P$ is known and incorporating in the analysis the new 
ETMC set of data that are coming out at a finer lattice spacing ($\beta=4.2$). 

Needless to say, the method can be straightforwardly extended to LQCD computations with
$u$, $d$, $s$ and possibly $c$ dynamical quarks where quenching uncertainties are virtually 
absent. In this respect we wish to note that in simulations with $N_f=3$ dynamical quarks, although low
energy hadronic effects in the $B$-meson wave function are correctly treated, 
a conflict remains between the number ($N_f=3$) of dynamical quarks running in the loops (and thus
relevant for the subtraction of UV divergencies) and the number ($N_f=4$) that instead should be used for
continuum RG-evolution at scales above, say, 1.5~GeV. This problem and the related RG-uncertainties 
are completely removed if also the $c$ quark is made dynamical.

Finally, we remark that the strategy we have outlined 
can be applied to any other $h\ell$ physical quantity the large $\mu_h$ 
behaviour of which is known (typically from large quark mass arguments). 

\vspace{0.5cm}
{\bf{Acknowledgments - }} We wish to thank R. Sommer and N. Tantalo for useful 
discussions and all the other members of ETMC for their interest in this work 
and a most enjoyable and fruitful collaboration. 

\vspace{1.2cm} 
\appendix

\renewcommand{\thesection}{A}
\section*{Appendix A -- Chiral and continuum extrapolations}
\label{sec:APP}   

For the reader's convenience we collect in this Appendix the standard formulae
we have used to perform the necessary chiral ($\hat\mu_l \to \hat\mu_{u/d}$)
and continuum  ($a\to 0$) extrapolations of our lattice data on pseudoscalar
masses and decay constants.

\subsection*{Fit ansatz for $h\ell$ meson masses and decay constants}
\label{FAMD}

We have modeled the $\hat\mu_\ell$-dependence of $M_{h\ell}$ and $f_{h\ell}$ in a form that is 
consistent with the known results of (NLO) SU(2) chiral effective theories for pseudoscalar mesons made 
up by a light plus a heavy quark. The form of the fit ansatz was chosen generic enough so as to
encompass (see discussion below) the expected $\hat\mu_\ell$-dependence both in the case when
the heavy quark is treated as static~\cite{GOI,GRI,SHZ} and when the latter is considered 
non-light but still relativistic~\cite{ROE,ALL}. 

In $r_0$-units we write for masses and decay constants
\beq
M_{h\ell} r_0 = C_0 + C_1 \hat\mu_\ell r_0 + \frac{a^2}{r_0^2} C_L \, ,
\label{MCC}
\eeq
\beq
f_{h\ell} r_0 = D_0 + D_1 \hat\mu_\ell r_0 + d_1
\frac{2B_0\hat\mu_\ell }{ (4\pi f_0)^2 } \log \left(
\frac{2B_0\hat\mu_\ell  }{ (4\pi f_0)^2 } \right) +
\frac{a^2}{r_0^2} D_L \, ,
\label{FCC}
\eeq
where the last terms, with parameters $C_L$ and $D_L$, have been included to cope with the 
expected O($a^2$) discretization effects. The fit parameters $C_{...}$'s and $D_{...}$'s in general 
depend on $\hat\mu_h$, though the form of the fit ansatz~(\ref{MCC}) and~(\ref{FCC}) 
has been actually employed only for a fixed value of the heavy quark mass, namely for 
$\hat\mu_h = \hat\mu_h^{(1)}$, when we evaluate the so-called triggering meson mass and decay 
constant, respectively. 

We checked that, within the statistical accuracy of our data, no $\hat\mu_\ell$ dependence 
is visible in the O($a^2$) terms in eqs.~(\ref{MCC}) and~(\ref{FCC}). The fit ansatz for 
$M_{h\ell}r_0$ does not include logarithmic terms. This is consistent with the results of the 
chiral effective theory for $h\ell$ pseudoscalar mesons with a light plus a non-light and 
relativistic quark~\cite{ROE,ALL}, but it can be equally well regarded as a simple Taylor 
expansion leading to a polynomial interpolation of data points with a very smooth 
dependence on $\hat\mu_\ell$ (see fig.~\ref{fig:fitm}). In this sense the fit ansatz for 
$M_{h\ell}r_0$ is also consistent with the spirit of the effective theory for static-light 
mesons~\cite{GOI,GRI,SHZ}, where the $\hat\mu_\ell$ dependence of $M_{h\ell}$ is 
expected to be a tiny effect (as we indeed find). 

The coefficient $d_1$, multiplying the term $\sim \hat\mu_\ell \log(\hat\mu_\ell)$ in 
eq.~(\ref{FCC}), was taken as a free fit parameter. Numerically we get for  
$d_1/D_0$ at $\hat\mu_h = \hat\mu^{(1)}_h$ a value ($-1.0 \pm 0.4$) which 
falls in between (and agrees within statistical errors with) what is expected from the 
arguments of refs.~\cite{GOI,GRI,SHZ} (where the heavy quark is treated as a static 
source) and those of ref.~\cite{ROE} (where it is taken as a relativistic particle).
The result we find for $d_1$ is not surprising as our heavy quark mass lies in the charm region.

The low energy constant $f_0$ and $B_0$ have been taken from recent ETMC
analyses of light meson quantities~\cite{DFHUW_Latt07}.

\subsection*{Fit ansatz for ratios}
\label{FAR}

For the quantities 
\beq
\frac{M_{h\ell}^L(\hat\mu_h^{(n)};\hat\mu_\ell,a)}
{M_{h\ell}^L(\hat\mu_h^{(n-1)};\hat\mu_\ell,a)}\, , \qquad 
\frac{f_{h\ell}^L(\hat\mu_h^{(n)};\hat\mu_\ell,a)}
{f_{h\ell}^L(\hat\mu_h^{(n-1)};\hat\mu_\ell,a)} \, ,  \quad n=2,3,4\, ,
\label{mfratfit}
\eeq
which enter the ratios $y$ and $z$ at the various $\hat\mu_h$-values, 
we employed fit ansatz analogous to (and derived from) eqs.~(\ref{MCC})
and~(\ref{FCC}) above. They all are of the following form: a leading term plus a
term linear in $\hat\mu_\ell$ and another one proportional to $a^2$ (and
$\hat\mu_\ell$-independent). 

Note that, if we assume that $d_1$ does not appreciably vary as $\hat\mu_h$ changes 
by the factor $\lambda \sim$~1.27--1.28 (which is a natural expectation in any effective 
theory for $h\ell$ mesons), one finds that in the ratios~(\ref{mfratfit}) the possible 
$\hat\mu_\ell \log(\hat\mu_\ell)$ dependence cancels at NLO, and is pushed to NNLO.


\vspace{0.3cm}
\begin{thebibliography}{99}

\bibitem{REVSM}
C. Amsler {\em et al.} [Particle Data Group], Physics Letters B667, 1 (2008)
and 2009 partial update for the 2010 edition.

\bibitem{UT}
http://www.utfit.org/  \\
http://www.slac.stanford.edu/xorg/ckmfitter/

\bibitem{REVLAT}
  K. Jansen,
  %``Lattice QCD: a critical status report,''
  arXiv:0810.5634 [hep-lat];\\
  %%CITATION = ARXIV:0810.5634;%%
  A.S. Kronfeld,
  %``The weight of the world is quantum chromodynamics,''
  Science {\bf 322} (2008) 1198;\\
  %%CITATION = SCIEA,322,1198;%%
  S. Durr {\it et al.},
  %``Ab Initio Determination of Light Hadron Masses,''
  Science {\bf 322}, 1224 (2008).
  %[arXiv:0906.3599 [hep-lat]].
  %%CITATION = SCIEA,322,1224;%%

\bibitem{HQET}
  E. Eichten and B. Hill, Phys. Lett. B {\bf 234} (1990) 511; \\
  H. Georgi,  Phys. Lett. B {\bf 240} (1990) 447;\\
  N. Isgur and M.B. Wise, Phys. Lett. B {\bf 232} (1989) 113 and Phys. Lett. B {\bf 237} (1990) 527;\\ 
  G.S. Bali,
  %``QCD forces and heavy quark bound states,''
  Phys. Rept. {\bf 343} (2001) 1.
  %arXiv:hep-ph/0001312];\\
  %%CITATION = PRPLC,343,1;%%

\bibitem{HQET1}
  M. Neubert,
  %``Heavy quark symmetry,''
  Phys. Rept. {\bf 245} (1994) 259.
  %[arXiv:hep-ph/9306320];\\
  %%CITATION = PRPLC,245,259;%%

\bibitem{SOMMER}
%Matching HQET to QCD: hep-lat/0310035, JHEP 02 (2004) 022
  J. Heitger and R. Sommer  [ALPHA Collaboration],
  %``Non-perturbative heavy quark effective theory,''
  JHEP {\bf 0402} (2004) 022;\\
  %[arXiv:hep-lat/0310035];\\
  %%CITATION = JHEPA,0402,022;%%
J. Heitger {\it et al.} [ALPHA Collaboration] JHEP {\bf 0411} (2004) 048;\\
%e-Print: hep-ph/0407227
 R. Sommer,
  %``Non-perturbative QCD: Renormalization, O(a)-improvement and matching to
  %heavy quark effective theory,''
  arXiv:hep-lat/0611020;\\
  %%CITATION = HEP-LAT/0611020;%%
  M. Della Morte, S. Durr, J. Heitger, H. Molke, J. Rolf, A. Shindler and R. Sommer
                  [ALPHA Collaboration],
  %``Lattice HQET with exponentially improved statistical precision,''
  Phys. Lett. B {\bf 581} (2004) 93 [Erratum-ibid.\  B {\bf 612} (2005) 313]; \\
  %[arXiv:hep-lat/0307021].
  %%CITATION = PHLTA,B581,93;%%
% some QUENCHED applications of the method to computing f_B, f_Bs and m_b
M. Della Morte, N. Garron, M. Papinutto and R. Sommer,
  %``Heavy quark effective theory computation of the mass of the bottom quark,''
  JHEP {\bf 0701} (2007) 007;\\
  %[arXiv:hep-ph/0609294].
  %%CITATION = JHEPA,0701,007;%%
M. Della Morte, S. Durr, D. Guazzini, R. Sommer, J. Heitger and A. Juttner,
  %``Heavy-strange meson decay constants in the continuum limit of quenched
  %QCD,''
  JHEP {\bf 0802} (2008) 078.
  %[arXiv:0710.2201 [hep-lat]].

\bibitem{FSSM}
 M. Guagnelli, F. Palombi, R. Petronzio and N. Tantalo,
  %``f(B) and two scales problems in lattice QCD,''
  Phys. Lett. B {\bf 546}, 237 (2002);\\
  %[arXiv:hep-lat/0206023];\\
  %%CITATION = PHLTA,B546,237;%%
  G.M. de Divitiis, M. Guagnelli, R. Petronzio, N. Tantalo and F. Palombi,
  %``Heavy quark masses in the continuum limit of quenched Lattice QCD,''
  Nucl. Phys.  B {\bf 675} (2003) 309;\\
  %[arXiv:hep-lat/0305018];\\
  %%CITATION = NUPHA,B675,309;%%
  G.M. de Divitiis, M. Guagnelli, F. Palombi, R. Petronzio and N. Tantalo,
  %``Heavy-light decay constants in the continuum limit of lattice QCD,''
  Nucl.\ Phys.\  B {\bf 672} (2003) 372.
  %[arXiv:hep-lat/0307005].
  %%CITATION = NUPHA,B672,372;%%
% application QUENCHED to f_Bs ...
D. Guazzini, R. Sommer and N. Tantalo,
  %``Precision for B-meson matrix elements,''
  JHEP {\bf 0801} (2008) 076.
  %[arXiv:0710.2229 [hep-lat]].
  %%CITATION = JHEPA,0801,076;%%

\bibitem{FERMILAB}
A.X. El-Khadra, A.S. Kronfeld and P.B. Mackenzie,
  %``Massive Fermions in Lattice Gauge Theory,''
  Phys. Rev.  D {\bf 55} (1997) 3933;\\
  %[arXiv:hep-lat/9604004];\\
  %%CITATION = PHRVA,D55,3933;%%
%
   S. Aoki, Y. Kuramashi and S. Tominaga,
   %``Relativistic heavy quarks on the lattice,''
   Prog.\ Theor.\ Phys.\  {\bf 109} (2003) 383; \\
   %[arXiv:hep-lat/0107009].
   %%CITATION = PTPKA,109,383;%%
%
   N.H. Christ, M. Li and H.W. Lin,
   %``Relativistic heavy quark effective action,''
   Phys. Rev.  D {\bf 76} (2007) 074505; \\
   %[arXiv:hep-lat/0608006].
   %%CITATION = PHRVA,D76,074505;%%
%
M.B. Oktay and A.S. Kronfeld,
  %``New lattice action for heavy quarks,''
  Phys. Rev.  D {\bf 78} (2008) 014504.
  %[arXiv:0803.0523 [hep-lat]].
  %%CITATION = PHRVA,D78,014504;%%

\bibitem{GAMIZLAT08}
  E. Gamiz,
  %``Heavy flavour phenomenology from lattice QCD,''
  arXiv:0811.4146 [hep-lat].
  %%CITATION = ARXIV:0811.4146;%%

\bibitem{ETMCD}
Ph. Boucaud {\em et al.} [ETM Collaboration], Phys. Lett. B {\bf 650} (2007) 304;\\ %hep-lat/0701012.
Ph. Boucaud {\em et al.} [ETM Collaboration],
  %``Dynamical Twisted Mass Fermions with Light Quarks: Simulation and Analysis
  %Details,''
  Comput. Phys. Commun. {\bf 179} (2008) 695;\\
  %arXiv:0803.0224 [hep-lat].
  %%CITATION = ARXIV:0803.0224;%%
B. Blossier {\em et al.} [ETM Collaboration], JHEP {\bf 0804} (2008) 020.
%
%%%!!!
%%%  B. Blossier {\em et al.},
%%%    %``Pseudoscalar decay constants of kaon and D-mesons from Nf=2 twisted mass
%%%    %Lattice QCD,''
%%%    JHEP {\bf 0907} (2009) 043.
%%%    %[arXiv:0904.0954 [hep-lat]].
%%%    %%CITATION = JHEPA,0907,043;%%  added, first ETMC paper with data at 3 beta's

\bibitem{VCS}
  B. Blossier {\it et al.} [ETM Collaboration], JHEP {\bf 0907} (2009) 043.
  %``Pseudoscalar decay constants of kaon and D-mesons from Nf=2 twisted mass
  %Lattice QCD,''
  %arXiv:0904.0954 [hep-lat].
  %%CITATION = ARXIV:0904.0954;%%

\bibitem{TM}  
R. Frezzotti, P.A. Grassi, S. Sint and P. Weisz, 
%Nucl. Phys. {\bf B} (Proc. Suppl.) {\bf 83} (2000) 941 and 
JHEP {\bf 0108} (2001) 058; \\
R. Frezzotti, S. Sint and P. Weisz [ALPHA Collaboration], JHEP {\bf 0107} 
(2001) 048; \\%(hep-lat/0104014);\\ 
M. Della Morte, R. Frezzotti, J. Heitger and S. Sint [ALPHA Collaboration], 
JHEP {\bf 0110} (2001) 041.\\
For a review, see A. Shindler, Phys. Rep. {\bf 461} (2008) 37.

\bibitem{LITERUQ}
  V. Gimenez, L. Giusti, G. Martinelli and F. Rapuano,
  %``NNLO unquenched calculation of the b quark mass,''
  JHEP {\bf 0003} (2000) 018; \\
  %[arXiv:hep-lat/0002007].
  %%CITATION = JHEPA,0003,018;%%
%
%\cite{Di Renzo:2004xn}
  F. Di Renzo and L. Scorzato,
  %``The N(f) = 2 residual mass in perturbative lattice-HQET for an
  %improved determination of the m(b)(MS-bar)(m(b)(MS-bar)),''
  JHEP {\bf 0411} (2004) 036;\\
%  [arXiv:hep-lat/0408015].
  %%CITATION = JHEPA,0411,036;%% 
C. Mc Neile {\it et al.} [UKQCD Collaboration], 
Phys. Lett. B {\bf 600} (2004) 77; \\
%
% Gray et al @ 2005: m_b and f_B with Nf=2+1
  A. Gray, I. Allison, C.T.H. Davies, E. Dalgic, G.P. Lepage,
J. Shigemitsu and M. Wingate,
  %``The Upsilon spectrum and m_b from full lattice QCD,''
  Phys. Rev.  D {\bf 72} (2005) 094507; \\
  %[arXiv:hep-lat/0507013].
  %%CITATION = PHRVA,D72,094507;%%
A. Gray {\em et al.}  [HPQCD Collaboration],
  %``The B Meson Decay Constant from Unquenched Lattice QCD,''
  Phys. Rev. Lett.  {\bf 95} (2005) 212001; \\
  %[arXiv:hep-lat/0507015].
  %%CITATION = PRLTA,95,212001;%%
%
%
E. Gamiz, C.T.H. Davies, G.P. Lepage, J. Shigemitsu and M. Wingate
                  [HPQCD Collaboration],
%``Neutral $B$ Meson Mixing in Unquenched Lattice QCD,''
Phys. Rev. D {\bf 80} (2009) 014503;\\
%  arXiv:0902.1815 [hep-lat].
%%CITATION = ARXIV:0902.1815;%%
   C. Bernard {\em et al.},
   %``B and D Meson Decay Constants,''
   PoS {\bf LATTICE2008} (2008) 278.
   %[arXiv:0904.1895 [hep-lat]].
   %%CITATION = POSCI,LATTICE2008,278;%%

\bibitem{FR2}
R. Frezzotti and G.C. Rossi, JHEP {\bf 0410} (2004) 070.

\bibitem{CR}
K.G. Chetyrkin, Phys. Lett. B {\bf 404} (1997) 161; \\
J.A. Vermaseren, S.A. Larin and T. van Ritbergen, Phys. Lett. B {\bf 405} (1997) 327; \\
K.G. Chetyrkin and A. Retey, Nucl. Phys. B {\bf 583} (2000) 3.
%e-Print: hep-ph/9910332

\bibitem{POLEtoMSbarMASS}
N. Gray {\em et al}, Z. Phys. C {\bf 48} (1990) 673; \\
D.J. Broadhurst, N. Gray and K. Schilcher,  Z. Phys. C {\bf 52} (1991) 111; \\
K.G. Chetyrkin and M. Steinhauser, Phys. Rev. Lett. {\bf 83} (1999) 4001; \\
K. Melnikov and T. van Ritbergen, Phys. Lett. B {\bf 482} (2000) 99.

\bibitem{DFHUW_Latt07}
P. Dimopoulos {\it et al.} [ETM Collaboration], PoS(LATTICE 2007) 102; \\
R. Baron {\it et al.} [ETM Collaboration],
  %``Light Meson Physics from Maximally Twisted Mass Lattice QCD,''
  arXiv:0911.5061 [hep-lat].
  %%CITATION = ARXIV:0911.5061;%%

\bibitem{Latt08}
P. Dimopoulos {\it et al.} [ETM Collaboration], arXiv:0810.2873 [hep-lat] and in preparation. 
%; \\ ETMC internal notes and in preparation.
%% we mean the scaling paper, see e.g. presentation by GH-CU-FF at Autrans meeting

\bibitem{FUT}
ETM Collaboration, in preparation. 

\bibitem{GL}
J. Gasser and H. Leutwyler. Phys. Rept. {\bf 87} (1982) 77, Ann. Phys. {\bf 158} (1984) 142 and Nucl. Phys. B {\bf 250} (1985) 465;\\
S. Capitani, M. L\"uscher, R. Sommer and H. Wittig  [ALPHA Collaboration], Nucl. Phys. B {\bf 544} (1999) 669. 

\bibitem{MarSac96}
G. Martinelli and C.T. Sachrajda,
  %``On the difficulty of computing higher-twist corrections,''
  Nucl.\ Phys.\  B {\bf 478} (1996) 660.
 % [arXiv:hep-ph/9605336].
  %%CITATION = NUPHA,B478,660;%%

\bibitem{RS}
J. Rolf and S. Sint [ALPHA Collaboration], JHEP {\bf 0212} (2002) 007. 
%hep-ph/0209255

\bibitem{DELM}
M. Della Morte, N. Garron, M. Papinutto and R. Sommer, JHEP {\bf 0701} (2007) 007. 
%hep-ph/0609294)

\bibitem{CHENEW}
K.G. Chetyrkin, J.H. Kuhn, A. Maier, P. Maierhofer, P. Marquard, M. Steinhauser and C. Sturm, arXiv:0907.2110 [hep-ph].

\bibitem{ZAPT}
  K.G. Chetyrkin and A.G. Grozin,
  %``Three-loop anomalous dimension of the heavy-light quark current in HQET,''
  Nucl. Phys. B {\bf 666} (2003) 289.
  %[arXiv:hep-ph/0303113].
  %%CITATION = NUPHA,B666,289;%%

%%%!!!  \bibitem{VCS}
%%%  %$f_K/f_\pi$ and $f_{D,D_s}$ paper  
%%%    B. Blossier {\it et al.} [ETM Collaboration], JHEP {\bf 0907} (2009) 043. 
%%%    %``Pseudoscalar decay constants of kaon and D-mesons from Nf=2 twisted mass
%%%    %Lattice QCD,''
%%%    %arXiv:0904.0954 [hep-lat].
%%%    %%CITATION = ARXIV:0904.0954;%%

\bibitem{ETMC_m_strange} 
ETM Collaboration, in preparation. 

\bibitem{CLEO_jan09}
J.P. Alexander {\it et al.}  [CLEO Collaboration],
  %``Measurement of $B{D_s^+ \to \ell^+ \nu}$ and the Decay Constant $fD_s^+$
  %From 600 $/pb^{-1}$ of $e^\pm$ Annihilation Data Near 4170 MeV,''
  Phys.\ Rev.\  D {\bf 79} (2009) 052001.
  %[arXiv:0901.1216 [hep-ex]].
  %%CITATION = PHRVA,D79,052001;%%

\bibitem{BECIR98}
  D. Becirevic, P. Boucaud, J.P. Leroy, V. Lubicz, G. Martinelli, F. Mescia and F. Rapuano,
  %``Non-perturbatively improved heavy-light mesons: Masses and decay
  %constants,''
  Phys. Rev.  D {\bf 60} (1999) 074501.
  %[arXiv:hep-lat/9811003].
  %%CITATION = PHRVA,D60,074501;%%

\bibitem{FR1}
R. Frezzotti and G.C. Rossi, JHEP {\bf 0408} (2004) 007. %(hep-lat/0306014)

\bibitem{GOI}
J.L. Goity, Phys. Rev. D {\bf 46} (1992) 3929. 

\bibitem{GRI}
B. Grinstein, E.E. Jenkins, A.V. Manohar, M.J. Savage, M.B. Wise, Nucl. Phys. B {\bf 380} (1992) 369.

\bibitem{SHZ}
S.R. Sharpe, Y. Zhang, Phys. Rev. D{\bf 53} (1996) 5125.

\bibitem{ROE}
A. Roessl,
  %``Pion kaon scattering near the threshold in chiral SU(2) perturbation
  %theory,''
  Nucl. Phys.  B {\bf 555} (1999) 507.
%  [arXiv:hep-ph/9904230].
  %%CITATION = NUPHA,B555,507;%%

\bibitem{ALL}
C. Allton {\it et al.}
[RBC-UKQCD Collaboration],
  %``Physical Results from 2+1 Flavor Domain Wall QCD and SU(2) Chiral
  %Perturbation Theory,''
  Phys. Rev.  D {\bf 78} (2008) 114509.
 % [arXiv:0804.0473 [hep-lat]]
  %%CITATION = PHRVA,D78,114509;%%


\end{thebibliography}
\end{document}